# Graph-Based Method for Anomaly Prediction In Brain Networks

Jalal Mirakhorli[1]     jalalmiry@aut.ac.ir
Hamidreza Amindavar[1]     hamidami@aut.ac.ir
Mojgan Mirakhorli[2]     genomic66@gmail.com

[1]Department of Electrical Engineering, Amirkabir University of Technology.
[2]Medical Genetic Lab, Iranian Comprehensive Hemophilia Care Center (ICHCC).

## Abstract:

Functional magnetic resonance imaging (fMRI) in neuroimaging techniques have improved in brain disorders, dysfunction studies via mapping the topology of the brain connections, i.e. connectopic mapping. Since, there are the slight differences between healthy and unhealthy brain regions and functions, investigation into the complex topology of functional and structural brain networks in human is a complicated task with the growth of evaluation criteria. Irregular graph deep learning applications have widely spread to understanding human cognitive functions that are linked to gene expression and related distributed spatial patterns, because the neuronal networks of the brain can hold dynamically a variety of brain solutions with different activity patterns and functional connectivity, these applications might also be involved with both node-centric and graph-centric tasks. In this paper, we performed a novel approach of individual generative model and high order graph analysis for the region of interest recognition areas of the brain which do not have a normal connection during applying certain tasks and resting-state or decompose irregular observations. Here, we proposed a high order framework of Graph Auto-Encoder (GAE) with a hypersphere distributer for functional data analysis in brain imaging studies that is underlying non-Euclidean structure in the learning of strong non-rigid graphs among large scale data. In addition, we distinguished the possible modes of correlations in abnormal brain connections. Our finding will show the degree of correlation between the affected regions and their simultaneous occurrence over time that can be used to diagnose brain diseases or revealing the ability of the nervous system to modify in brain topology at all angles, brain plasticity, according to input stimuli.

## Introduction:

The human brain has a complex connection of various parts which dynamically shift during their operation. Therefore, the model and cost of each part is able to change according to the type of its operation in carried out or rest state. The fMRI data exhibits non-stationary properties in the context of task-based studies (Hutchison et al. 2013; Calhoun et al. 2014), the analysis of these sections is known to predict the connection factors for each independent profile. Here, we present a theoretical model based on high order VAE and graph theory to learn the probability distribution of the graph that is known to extract the data model of tasks from brain regions with a semi-unknown prior knowledge method. We used each functional connectivity matrix that is collected in a resting-state fMRI (rs-fMRI) experiment, using rs-fMRI data from Alzheimer's Disease Neuroimaging Initiative (ADNI) dataset. Functional connectome analysis is recognized to reveal biomarkers of individual psychological or clinical traits and describes the pairwise statistical dependencies between brain regions. In this article, we present the brain as a graph by means of functional connectome structures. This allows us to probing and infer, how dynamic changes progress of improvement degree in brain disorder or predict the disease as well as identify the term brain abnormalities. This paper proposes to introduce a framework for feature extraction of the brain graphs which provide across many subjects, for prediction of ambiguous parts of the brain. In this method a Variational Autoencoder (VAE) is established to make the graph and experiment a Bayesian Von Mises–Fisher (VMF) (Mardia et al. 1976) mixture model as a latent distribution that can place mass on the surface of the unit hypersphere (Banerjee et al. 2005) and stable the VAE. Our experiments demonstrate that this method significantly outperforms other methods and is a large step forward to infer brain structure. It is capable to handle both homogeneous and heterogeneous graphs. According to the recent studies, the geometric deep learning methods have been successfully applied to data residing on graphs and manifolds in terms of various tasks (Bronstein et al. 2017; Mirakhorli et al .2017). For example, function of the brain in predicting and its graph expression analysis address the multifaceted challenges arising in diagnosis of brain diseases. Here, we present a novel method using a high order graph model in revealing the relationship between the parts of brain and recover missing parts or malfunctioning parts. The method can also be predicted the effects of long-term deep brain stimulation on brain structural and functional connectivity.

**Related works:**

As our approach focuses on completing the graph and predictive defective parts of the graph via obtained feature of network embedding, we review some of the state-of-the-art research that are close to our work. Xu et al. 2017 constructed a graph from a set of object proposals to provide initial embedding to each node and edge while using message passing to obtain a consistent prediction. Simonovsky et al. 2018 used a generative model to produce a probabilistic graph from a single opaque vector without specifying the number of nodes or the structure explicitly. Pan et al. 2018 proposed an adversarial training scheme to regularize and enforce the latent code to match a prior distribution with a graph convolutional Autoencoder. Makhzani et al. 2015 showed an adversarial Autoencoder to learn the latent embedding by merging the adversarial mechanism into Autoencoder for general data. However Dai et al. 2017 applied the adversarial procedure for the graph embedding. Also an encoder with edge condition convolution (ECC) (Johnson et al. 2017) was used for conditioning both encoder and decoder which was associated with each of the input graphs (Simonovsky at al. 2018), this method is useful only for generation small graphs. In addition, we used a combination of graph convolution VAE to address both recovery and learning problems which can perform in spectral (Defferrard et al. 2016; Levie et al. 2019) or spatial domains (Monti et al. 2017).

**Materials and Method**

In spite of individual alteration, the human brains perform common patterns among different subjects. Therefore, algorithms based on graphs are essential tools to capture and model the complicated relationship between functional connectivity. In this work, we used a model of graph embedding to convert graph data into a low dimensional and continuous compaction feature space that is able to detect abnormal parts of input graphs which is involved with graph matching and partial graph completion problems (Verma et al. 2018). To develop this algorithms, we need to present a generative model that is constructed from a high order Graph Variational Autoencoder with hypersphere distribution (Davidson et al. 2018; Kingma et al. 2014; Kipf et al. 2016). Partial abnormality is able to be demonstrated by features train in latent space, considering both first-order proximity, the local pairwise proximity between the vertices in the network, and second-order proximity. This refers to vertices sharing many connections to other vertices that are similar to each other. The work flow of the algorithm is shown in figure 1.

Brain network as a graph: As shown in figure 1, using rs-fMRI data of subjects acquired by preprocessing ADNI dataset to provide an adjacency matrix that encodes similarities between nodes and a feature matrix that represents a node's connectivity profile, to define the input data as an undirected graph. In this paper, we define the connected graph G = (V, E, W), which consists of a finite set of vertices V with |V| = n, a set of edges E, and a weighted adjacency matrix W. If there is an edge e = ( i, j) connecting vertices i and j, the entry $W_{ij}$ or $a_{ij}$ represent the weight of the edge $a_{ij>0}$, otherwise $a_{ij} = 0$. For each of n subjects make a data matrix $X_n \varepsilon R^{dn*dy}$, where $d_y$ is the dimension of the node's feature vector. This structure of fMRI data will be merge into the graph defined. We will show that applying the graph based algorithm on brain connectivity is useful to analyze brain information processing.

Graph Convolutional Neural Network: For applying convolution-like operators over irregular local supports, as graphs where nodes can have a varying number of neighbors which can be used as layers in deep networks, for node classification or recommendation, link prediction and etc. This process is involved with three challenges, a) defining translation structure on graphs to allow parameters sharing, b) designing compactly supported filters on graphs, c) aggregating multi-scale information, the proposed strategies broadly fall into two domains, there is one spatial operation which directly performs the convolution by aggregating the neighbor nodes' information in a certain batch of the graph, where weights can be easily shared across different structures (Niepert et al. 2016; Gao et al.2018) and the second is a spectral operation which relies on the Eigen-decomposition of the Laplacian matrix that is applied to the whole graph at the same time(Henaff et al. 2015; Levie et al. 2019; Bruna et al. 2014; Kipf at al. 2017), spectral-based decomposition is often unstable making the generalization across different graphs difficult (Pan et al. 2018), that cannot preserve both the local and global network structures which require large memory and computation. On the other hand, local filtering approaches (Boscaini at al. 2016) rely on possibly suboptimal hard-coded local pseudo-coordinates over graph to define filters. The third approach relies on point-cloud representation (Klokov et al. 2017) that cannot leverage surface information encoded in meshes or need ad-hoc transformation of mesh data to map it to the unit sphere (Sinha et al. 2016). Overall, the spectral approach has the limitation of graph structure being the same for all samples i.e. homogeneous structure, this is a hard constraint, as most of sample graphs in the learning phase have the same structure and size for different samples i.e. heterogeneous structures. Therefore, we applied the spatial approach that is not obligatory homogeneous graph structure, but in turn requires preprocessing of graph to enable learning on it and used a method that proposes a graph embed pooling. Graph convolution transforms only the vertex values (Such et al. 2017) whereas graph pooling transforms both the vertex values and the adjacency matrix. Convolution of vertices V with filter H only requires matrix multiplication of the form, $v_{out}=Hv_{in}$ where $v_{in}$, $v_{out}$ $\varepsilon\ R^{N*N}$. the filter H is defined as the k-th degree polynomial of the graph adjacency matrix A;

$$H=h_0I+h_1A+h_2A^2+\ldots+h_nA^k, H\ \varepsilon\ R^{N*N}. \quad (1)$$

We used the first three taps of H for any given filter.

Graph Autoencoder (GAE): GAE is inherently an unsupervised generative model, our model is almost based on the framework of VAE that was produced in (Kipf et al. 2016; Xu et al. 2018). In follow, we briefly describe GAE and introduce our method with objectives. For learning both encoder, decoder in the figure 1 to map between the space of graph and their continuous embedding $Z\ \varepsilon\ R^C$, stochastic graph encoder $q_\Phi(Z|G)$ embed the graph into continuous representation Z. Given a point in the latent space Z, the graph decoder $p_\theta(G|Z)$ outputs a probabilistic fully-connected graph Ğ on predefined nodes, where $\Phi,\theta$ are learned parameters. Reconstruction ability of GAE is facilitated by approximate graph matching for aligning G with Ğ, as well as a prior distribution P(Z) imposed on the latent code representation as a regularization and train GAE via optimization of the marginal likelihood, $P(G)=\int P\theta(Z)P(G|Z)\,dz$, then the marginal log likelihood can be written;

$$\log p_\theta(G) = \acute{K\!L}(q_\Phi(Z|G)\ \|\ p_\theta(Z|G)) + \pounds(\theta,\Phi;G). \quad (2)$$

$$\pounds(\theta,\Phi;G) = -\ \acute{K\!L}[q_\Phi(Z|G)\ \|\ p_\theta(Z) + \tilde{\mathbb{E}}_{q_\Phi(Z|G)}\log p_\theta(G|Z)]. \quad (3)$$

Where Kullback–Leibler ($\acute{K\!L}$) and $q_\Phi(Z|G)$ are a divergence term in loss function that encourages the variational posterior and a variational approximation to the posterior distribution, p (Z|G),

respectively. Here, we used a hyperspherical latent structure for parameterization of both prior and posterior, because one of important limitation in using Gaussian mixture is that ḰĹ term may encourage the posterior distribution of the latent variable to collapse in prior or tends to pull the model toward the prior, during approximation the prior, whereas in the VMF (Fisher et al.1953; Kanti et al. 1976) case there is not such pressure toward a single distribution convergence. Therefore a VMF distribution is more suitable for capturing data (Kipf et al. 2016), VMF distribution defines a probability density over points on a unit-sphere also the consequences of ignoring the underlying spherical manifold are rarely analyzed in parts due to computational challenges imposed by directional statistics.

Geometric deep learning: For graph generation, we employed the GAE to graph $G \varepsilon R^{n*m}$ under an unsupervised learning method, our goal is to learn an implicit generative mode that can predict abnormal sections in the graph, indeed, we are not confident that close links have similar features to detect invisible deformable and hidden angle of graphs. Our method almost inspired in previous studies (Larsen et al. 2015; Wu et al. 2016), in combination from the GAE and generative adversarial network (GAN) that decoder of GAE and generator of GAN have been a supportive role. Following the above mentioned items, we used the uniform distribution VMF(0,Ҡ =0) for our prior and approximate $p_\theta (Z | Ğ)$ with variational posterior $q_\Phi(Z|G) = VMF(Z; \mu, Ҡ)$, where μ is mean parameter and Ҡ is a constant, the variational distribution is associated with a prior distribution over the latent variables, our GAE loss combines the graph reconstruction $Ĺ_r = || Ğ – G ||_2$ encouraging concatenation both the encoder-decoder to be a nearly identical transformation, a regularization prior loss measured by the ḰĹ divergence, $Ĺ_p = D_{ḰĹ}(q(z|G) || P(Z))$ and a cross entropy loss $Ĺ_{2D-GAN}$ for GAN, $Ĺ_{G-GAN} = \log D(G) + \log(1-D(G(z)))$, where D is discriminator as a confidence D(G) of the whether a input graph G is real or synthetic (Levie et al. 2019). The total GAE+GAN loss is computed as $Ĺ = Ĺ_r + \lambda_1 Ĺ_p + \lambda_2 Ĺ_{G-GAN}$, where $\lambda_1$ and $\lambda_2$ are weights of ḰĹ divergence loss and reconstruction loss. As discussed in above, our desire to focus on graph completion for deformable object classes in brain connectome. Therefore, we used dynamic weight of filtering in each convolutional layer.

Partial graph completion: Once our model GAE&GAN has been trained, the encoder and the element of GAN are discarded away, so that the role of the decoder is only as a graph generator where the probabilistic latent space z acts as a base for finding the target graphs for the same graph prior. At inference, for each space of the latent vector z* may represent a few complete graph correspondence a latent vector, then partial graph or deformation graph in the input of the system makes a few complete graphs in the output, the higher deformation rate in input, the more of graph is generated. Each partial graph represents a partial adjacency matrix δ that can be applied to any graph Ğ generated by our model and to explore similarity between them, for finding best compatibility or a latent vector z* which can minimize differences between input and output graph, to provide more geometric insight on the problem. Process to measure similarities among elements of graphs with probing combination of dependences similar unary, pairwise or high-order (Le-Huu et al. 2017; Yan et al. 2018; Yu et al. 2016) as well as there are potentials between reference graph and their counterparts similar to previous study (Wang et al.2019; Wang F et al. 2018), that followed a function used for finding high order dissimilarity or deformation into convex optimization problem over a set of doubly stochastic matrices.

Graph recovery plan: As mentioned above, our goal is the optimal choice of a latent vector z* so that minimal dissimilarities exists of between the partial graph related to a disease brain, G, and the generated graph Ğ =dec (z), or min (Ğ, ζG).

Where ζ denote a non-rigid transformation, this procedure is performed over z* and ζ, alternatively. Minimizing the following function is our goal as an objective function;

$$\min j(p, \zeta) = \sum_{i,j} p_{ij} \, ||v_{\breve{G}} - \zeta(v_G)||^2 + \gamma_{(\zeta)} \,. \qquad (4)$$

Where γ is a regularization term of geometric transformation ζ: Ğ⟶G, p is a map for measuring the difference of graph attributes in a similar transformation domain. In each step of optimization a weight matrix measures the degree of deformation on the radial basis function method. Graph recovery is an ill-posed problem that has multiple plausible solutions while in this paper we limit the prediction space to only several structures of the graphs.

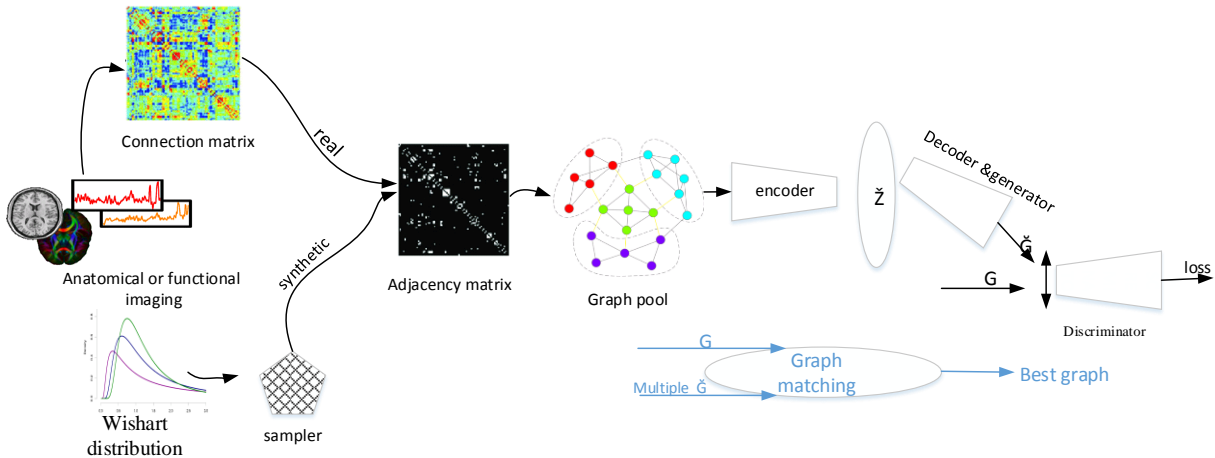

Fig. 1. The data flow of the proposed network architecture.

## Conclusion

Through this paper, we have presented a novel joint prediction approach for the selection of meaningful partial correlation and extract role of functional connectivity states from rs-fMRI to describe the dynamic connections. This method helps to improve biomarker discovery especially in high-dimensional settings with a large number of variables connections. This approach also allows us to recover data corrupted by noise and explore in domain of unknown function. We showed that the modeling latent space with the hypersphere distributer improve accuracy in prediction connectivity states of the brain. Based on the above analysis and focus on these topological attributes to extend this work, we will extract important information in the future on the higher-order function of the brain network via semantic functional rich-club organization.

**Acknowledgement**

The authors gratefully acknowledges the assistance provided by the Medical Genetic Lab, Iranian Comprehensive Hemophilia Care Center (ICHCC).